\documentstyle[preprint,aps]{revtex}
\begin{document}
\title
{Evidence of two-electron tunneling interference in Nb/InAs junctions}
\author{Antonio Badolato, Francesco Giazotto, Marco Lazzarino, Pasqualantonio Pingue, 
	and Fabio Beltram}
\address{Scuola Normale Superiore and Istituto Nazionale per la Fisica della Materia,
	I-56126 Pisa, Italy}
\author{Carlo Lucheroni and Rosario Fazio}
\address{Dip. di Metodologie Fisiche e Chimiche (DMFCI), Universit\`a di Catania and 
	Istituto Nazionale per la Fisica della Materia, I-95125 Catania,Italy }
\maketitle
\begin{abstract}
The impact of junction transparency in driving phase-coherent charge 
transfer across diffusive semiconductor-superconductor junctions is
demonstrated. We present conductivity data for a set of Nb-InAs junctions
differing only in interface transparency. Our experimental findings 
are analyzed within the quasi-classical Green-function approach
and unambiguously show the physical processes giving rise to
the observed excess zero-bias conductivity.
\end{abstract}
\vskip0.5cm
PACS numbers: 74.80.Fp, 73.23.Ad, 73.40.Qv

\newpage
The lack of available states below the superconducting energy gap 
prevents single-electrons from being injected across normal 
metal-superconductor (NS) junctions at low bias. A $two$-electron 
tunneling process 
known as {\it Andreev reflection} \cite{andr}, however, allows charge 
transfer across interfaces with non-zero transparency. The two 
electrons involved in this process upon entering the superconductor 
form a Cooper-pair and must therefore be linked by specific energy 
and phase relationships \cite{benRev,lambert}.

The detailed nature of the N region in proximity to the superconductor can have
a significant impact on junction conductivity \cite{vanwees}.
Indeed the interplay between phase-coherence in the N region and  Andreev
reflection can lead to enhanced or suppressed conductivity~\cite{kast91,reentrance}.
Finite bias, temperature, and magnetic fields 
readily destroy phase coherence and consequently these phenomena 
are observed mostly around zero bias and at very low temperatures. 

Experimentally, these coherent phenomena can be studied in the case of
diffusive N electrodes. 
Beenakker \cite{marmor} introduced the concept of
{\it reflectionless tunneling} where the zero bias conductivity
is predicted to present an unexpected linear dependence on
interface transmission probability (analogous to a one-electron process).
In the tunneling regime, Hekking and Nazarov  \cite{hn1,hn2} 
underlined the role of coherent phenomena in determining the constructive 
interference of the two electrons entering the superconductor. 
In the low transparency case, one can identify two regimes depending on the 
resistance of the disordered (phase coherent) region being larger or smaller 
than the barrier resistance. The excess conductivity around zero bias as a function of the latter increases or decreases, respectively.
Unified theoretical descriptions of these transport phenomena
are now available within the quasi-classical Green's function \cite{vzk}
and the scattering matrix \cite{lambert,benPrl} approaches.

The first case (interface resistance limiting the conductivity) was
invoked for a Cu-Al junction by Pothier {\it et al.} \cite{pot},
the second (N-region resistivity dominant) for the
superconductor-semiconductor junction results of Magn\'ee 
{\it et al.}\cite{magnee}.
Identification of the specific regime was obtained by analyzing
the relevant junction parameters.
In this article we take advantage of the qualitatively
different dependence of conductivity on junction transparency to demonstrate
two-electron tunneling interference in a set of InAs/Nb junctions. 
In our samples, in fact, zero-bias excess conductance develops for 
increasing the transparency of the junctions:
as far as we know this is the first direct experimental demonstration 
of this characteristic trend.
The data are quantitatively analyzed by solving the appropriate 
quasi-classical equations \cite{vzk}.

The samples consist of semiconductor-superconductor 
planar junctions, where the semiconductor part is a n-doped InAs 
(100) substrate 
with n\,=\,$1.3 \cdot 10^{18}$ cm$^{-3}$. The superconductor part is
Nb. The relevant semiconductor parameters are the resistivity $\rho=4 \cdot 10^{-4}
\,\Omega$ cm and the mobility $\mu=1\cdot 10^{4}\,$ cm$^2$ V$^{-1}$ s$^{-1}$.
InAs is one of the best choices for hybrid Sm-S devices, mostly
owing to the lack of a Schottky barrier at the contact with a metal. 
Oxides and other impurities, however, give rise to a residual interface 
$barrier$ region and several technological 
efforts have been made to improve the transparency of S-Sm junctions and
maximize Andreev conversion \cite{kast91,reentrance,tech}. In our case, 
junctions
were fabricated after annealing substrates at a moderate temperature 
(340 $^\circ$C) 
only marginally affecting the native oxide structure and its
nonuniformities. 
This annealing procedure, however, minimizes the amount of water incorporated by
the oxide and reduces the formation of Nb oxides\cite{neur}.
The superconductor film (100-nm-thick Nb) was e-beam evaporated following
substrate annealing and without breaking the vacuum.  
Substrate temperature during Nb deposition was kept at $\approx$150 $^\circ$C 
to promote adhesion of the superconducting film.

Circular junctions (75 $\mu$m diameter) were fabricated by standard 
photolithographic techniques and wet chemical
etching in a H$_2$O:H$_2$O$_2$:NaOH solution. Back contacting was
provided by metallizing the whole chip back.
Samples were measured from 0.3 K to temperatures larger than the critical 
temperature $T_c\simeq8.0$ K of our Nb-on-InAs film and for static magnetic 
fields applied perpendicularly to the plane of the junctions using 
a $^3$He closed-cycle cryostat equipped with a superconducting magnet.

Figure 1 shows the differential conductance $G(V)$ of several contacts
belonging to the same chip at $T= 0.32$ K with no applied magnetic field.
The wide range of conductance values observed is linked to the above
mentioned oxide inhomogeneity. Notably high conductivity contacts exhibit
zero-bias excess conductance. The latter effect was
consistently more pronounced in more conductive junctions.
As we shall argue in the following, more conductive junctions correspond
to more transparent interfaces. The qualitative behavior of our junctions is
therefore that
studied by Hekking and Nazarov and excess conductance
may be linked to the constructive interference of two-electron
tunneling into the superconductor. This conclusion will be substantiated 
by the following quantitative analysis.
High conductance curves are shown in a narrower range owing to
heating effects at higher biases, but did exhibit a peak in proximity
of $V=\Delta$/$e$ (data not shown). 

The experimental curves labeled {\it(1)} and {\it(2)} do not quite follow
the expected behavior for a SN junction.
The main discrepancies are in the detailed shape and in the position of 
the peaks at $\Delta$. These deviations are related to the complex structure of the 
samples in the interface region.
In order to describe electronic transport in our InAs/Nb junctions, we
adopted the model first proposed in Ref. \onlinecite{neur},
namely we take into account the fact that the first layers of Nb are 
non-superconducting owing to the fabrication process (see Fig. 2). 
The potential drop still 
occurs at the metallurgical interface and therefore this interlayer 
region ($N_{Nb}$) can
be considered at the same potential as the superconducting electrode. The
thickness of $N_{Nb}$ ($L$) is expected to be of the order of a nm; nevertheless
this interlayer strongly modifes the transport properties of the junctions. 
The various contributions to the current depend on the details of the 
sample close to the contact region and while the tunneling into the 
condensate (Andreev current) may not be very sensitive to the interlayer, 
the normal contribution strongly depends on the single-particle states 
available close to the interface~\cite{ldos}. 

The numerical evaluation of the current-voltage (I--V) characteristics is done in two steps.
First we determine self-consistently the BCS gap in the Nb (close to
the junction it will be suppressed due to proximity).
Since we assume that there is no potential drop in this part of the structure
we can use the equilibrium formalism. In the dirty limit, the quasiclassical retarded
Green's function $\hat{g}^{R}(x,E)$ satisfies the Usadel equation \cite{schmid81}
\begin{equation}
        D \; \partial _x (\hat{g}^{R} \cdot \partial_x \hat{g}^{R})
        + (iE + \Gamma_{in}(x))
        \left[\hat{\tau} _{z}+\hat{\Delta }, \hat{g}^{R} \right] 
        =0 ,
\label{usadel}
\end{equation}
with the constraints $\hat{g}^{R} \hat{g}^{R} = 1$ and $\mbox{Tr} \hat{g}^{R} = 0$.
The hat refers to the Nambu notation ($\hat{\tau}_z$ and $\hat{\tau}_y$
are the Pauli pseudospin matrices). 
The Fermi energy is at $E=0$.
We assume that the inelastic scattering rate $\Gamma_{in}(x)$ 
differs from zero only in the thin Nb interlayer close to the interface.
The gap matrix $\hat{\Delta}(x) = \Delta(x) \; \hat{\tau}_y\;$ is determined
self-consistently by means of
\begin{equation}
        \Delta(x) =\frac{1}{2}
        \lambda(x) \int^{\omega_D}_{0}dE\; {\rm Im} [g_{12}(x,E)
                   \tanh(E/2T) ] ,
        \; \;\; \;
\label{selfconsistent}
\end{equation}
where $\lambda(x)=\lambda \Theta (-x)$ ($\lambda= \mbox{arcosh}(\omega_D /
\Delta_{BCS})$ is the BCS coupling constant,
$\Theta(x)$ is the Heaviside step function, $\Delta_{BCS}$ is the BCS gap, $\omega_D$
is the Debye cutoff frequency) and $T$ is the temperature.
The selfconsistent solution close to the interface enables us to determine the
appropriate boundary condition for the determination of the I-V curves (along the
lines of Volkov et {\em al.}~\cite{vzk})
\begin{equation}
I = \frac{1}{eR_N}\int dE D(E) [f_0(E+eV/2) - f_0(E-eV/2)] ,
\end{equation}
where $R_N$ the normal state resistance of the sample and $f_0$ the Fermi distribution
function. 
The effective transmission coefficient of the structure
$$
	D(E) = \frac{1 + r}{rM^{-1}(E)+ \frac{1}{d}\int_0^d M_T^{-1}(x,E)dx}
$$
with 
$M(E)= \Re g_{11}^{R}(0^+)\Re g_{11}^{R}(0^-) +\Im g_{12}^{R}(0^+)\Im g_{12}^{R}(0^-)$  
and 
$M_T(x,E)=[\Re g_{11}^{R}(x,E)]^2 + [\Im g_{12}^{R}(x,E)]^2$
is determined by the solution of the Usadel equation. The interface between the 
diffusive region and $Nb$ is placed at $x=0$ ($x=0^{+/-}$ indicated the normal/
superconducting side of the interface).  
The result of this procedure leads to the fits shown in  Fig.~\ref{F3}.
The theoretical curves in Fig.~\ref{F3} are obtained assuming the 
ratio $r=R_{T}/R_{D}=2.5$ (where $R_T$ and $R_D$ are the interface
and diffusive region contribution to the junction resistance, respectively).
The fitting parameters
in the $N_{Nb}$ region are the length $L\simeq0.2\,\xi_{Nb}$ 
(with $\xi_{Nb}$ the experimental
coherence length in the dirty S) and the inelastic scattering rate
$\Gamma_{in} =0.4\,\Delta_{BCS}$. All the curves of the set are fitted 
with the same
parameters, and only the temperature is varied according to its experimental
value.

Both $L$ and $\Gamma_{in}$ are consistent with the experimental 
expectations. From the measured value of $T_c$ it was possible to 
extract 
\cite{vaglio} $\xi_{Nb}\simeq{8.0}$ nm for
our Nb-on-InAs film, obtaining in this way a value of the interlayer length
$L\simeq{1.6}$ nm consistent with the estimated thickness of the 
oxidized Nb interface region \cite{neur}. 
The large value of $\Gamma_{in}$ is related to the fact that
superconductivity is locally suppressed (we stress again that the rest 
of the Nb is ideal).
The presence of the thin layer of "normal" Nb proved crucial to obtain a 
good quantitative
agreement with the experimental data. 
In fact, at finite voltages the measured
differential conductance is
always larger than that predicted by a simple S-I-N system. 
This increased conductivity stems from the presence of the
$N_{Nb}$ layer and the availability of states for the
single-particle channel of the
current. 
The latter are the subgap states of the proximized normal layer. 

From the already given experimental parameters of InAs, we obtained a very 
good agreement between theory and experiments with  
$d=200$ nm (which corresponds roughly to the estimated coherence length in the normal region) and a mean free path $l_e=220$ nm at $T=0.3$ K \cite{z}. 
The fitting
parameter $r$ allows us to quantitatively show that for the junction
corresponding to curve {\it(1)} the weight of the tunnel-barrier
resistance is predominant with respect to the resistance of the 
diffusive region.
A quantitative fit for curve {\it (2)} was also obtained (data not shown). 
As anticipated above this occurred for a lower value of the parameter $r$ 
($r=2$ compared to $r=2.5$ for junction {\it (1)}, reflecting the larger 
transparency of the interface in the corresponding contact). Although the
various junctions are nominally identical, the native oxide nonuniformities 
can easily lead to the observed variations.
The parameter values extracted from the fits show that 
our junctions are in an intermediate case as compared to the limit discussed 
in Ref.~\onlinecite{hn1}. Though not deep in the asymptotic region, 
our analysis allows to extend the main qualitative features also to the 
range of parameters typical of the samples considered in this work.

The zero-bias conductance observed in the 
curves {\it(3)}, {\it(4)}, and {\it(5)} 
in Fig. 1 occurs on an increasing normal-state conductance value.
The latter variation is influenced also by a concomitant variation of the 
junction effective area \cite{area}. For the present analysis, however,
we must focus on the increased zero-bias contribution
which is dominated by the junction-transparency variation.
This corresponds to the Hekking-Nazarov regime, in which increasing
$\Gamma$ leads to a more pronounced conductance peak at $V,H=0$. 
In terms of our parameters this means that $r$ lowers. In fact,
this trend is also 
present in our model calculations: by lowering the barrier resistance 
the zero-bias excess conductance develops. For this set of curves, 
however, we were unable to obtain a
quantitative agreement with the experimental data.  

Further proof on the nature of the enhanced zero-bias conductivity can 
be gained experimentally 
by analyzing its temperature and magnetic field dependence. As mentioned above
two-electron interference effects can be easily broken by voltage, 
temperature and
magnetic field. Figure \ref{F4} shows the temperature and the magnetic-field
(inset) dependence of the zero-bias conductance in junction {\it(4)}. 
Cut-off values
for temperature ($T\simeq 1.0$ K) and magnetic field ($B\simeq 6$ mT) are similar to those
reported by other experimental groups for such effects \cite{kast91}. Additionally we remark that at $T=0.32$ K the 
width of
the zero bias peak is of the order of $k_{B}T/e=27$ $\mu$eV.

In conclusion, we have analyzed the influence of junction transparency
on phase-coherent conductance in superconductor-semiconductor
junctions. The dependence of the zero-bias excess
conductance on interface transparency was directly observed and 
junction parameters estimated based on a quasi-classical Green's
function approach. 
  
Very useful discussions with C. W. J. Beenakker and R. Raimondi are 
gratefully acknowledged. 
The present work was supported by INFM Sezione E 
under the PAIS project Eterostrutture 
Ibride Semiconduttore-Superconduttore.


\begin{figure}
\caption{ Experimental differential conductance vs voltage at $T=0.32$ K and
zero magnetic field for five Nb/InAs junctions belonging to the same chip.
Numbers in parentheses label different junctions.
The left and the right hand side insets show a magnification
of data for junctions {\it(3)}, {\it(4)}, {\it(5)}.
}
\label{F1}
\end{figure}

\begin{figure}
\caption{ Sketch of our one-dimensional model. S labels the pure Nb 
superconducting electrode, $N_{Nb}$ is a nonsuperconducting
but strongly proxymized portion of Nb of thickness $L$ in equilibrium with S. 
The solid vertical line represents the tunneling barrier contributing with 
$R_T$ to the junction resistance.
$N_D$ is the nonequilibrium portion of InAs of length $d$ and resistance $R_D$.
N labels the InAs electrode.
} 
\label{F2}
\end{figure}

\begin{figure}
\caption{ Normalized differential conductance vs voltage  with 
no applied magnetic field at three different temperatures 
$T=0.32,\,1.2,\,3.0$ K. Data refer to contact labeled 
{\it (1)} in Fig. 1. $R_N$ is 
the normal-state resistance.
Experimental: open triangles, theoretical: solid lines.
Parameters are $L=1.6$ nm, $d=200$ nm, $\Delta_{BCS}=1.265$ meV,
$\Gamma_{in}=0.4\,\Delta_{BCS}$, $r=2.5$.
All experimental curves are fitted with the same 
parameters, only the temperature is varied according to its experimental value.
}
\label{F3}
\end{figure}

\begin{figure}
\caption{ Temperature dependence of the zero-bias conductance relative to 
contact labeled {\it (4)} in Fig. 1. The inset shows the magnetic-field 
dependence for the same contact at $T=0.32$ K. Magnetic field was applied 
perpendicularly to the junction plane.
}
\label{F4}
\end{figure}


\begin{references}

\bibitem{andr} A. F. Andreev, Zh. Eksp. Teor. Fiz. {\bf 46}, 1823 (1964)
	[Sov.Phys. JETP {\bf 19}, 1228 (1964)].

\bibitem{benRev} C. W. J. Beenakker, Rev. Mod. Phys. {\bf 69}, 731 (1997).

\bibitem{lambert} C. J. Lambert and R. Raimondi, J. Phys. Condens. Matter
	{\bf 10}, 901 (1998).

\bibitem{vanwees} B. J. van Wees, P. de Vries, P. Magn\'{e}e, and T. M. Klapwijk,
	Phys Rev. Lett {\bf 69}, 510 (1992).

\bibitem{kast91} A. Kastalsky, A. W. Kleinsasser, L. H. Greene, R. Bhat,
F. P. Milliken, and J. P. Harbison, Phys. Rev. Lett. {\bf 67}, 3026
(1991); C. Nguyen, H. Kroemer, and E. L. Hu, Phys. Rev. Lett {\bf 69}, 2847 (1992).

\bibitem{reentrance}  P. Charlat, H. Courtois, Ph. Gandit, D. Mailly, A. F. Volkov,
	and B. Pannetier, Phys. Rev. Lett. {\bf 77}, 4950 (1996).

\bibitem{hn1} F. W. J. Hekking and Yu. V. Nazarov, Phys. Rev. Lett. {\bf 71}, 1625 (1993).

\bibitem{hn2} F. W. J. Hekking and Yu. V. Nazarov, Phys. Rev. B {\bf 49}, 6847 (1994).

\bibitem{marmor} C. W. J. Beenakker, Phys. Rev. B {\bf 46}, 12841 (1992); I. K.
	Marmorkos, C. W. J. Beenakker, and R. A. Jalabert, Phys. Rev. B {\bf 48}, 2811 (1993).

\bibitem{vzk} A. F. Volkov, A. V. Zaitsev, and T. M. Klapwijk, Physica C {\bf 210}, 21 (1993).

\bibitem{benPrl} C. W. J. Beenakker, B. Rejaei, and J. A. Melsen, Phys. Rev. Lett.
	{\bf 72}, 2470 (1994).

\bibitem{pot} H. Pothier, S. Gu\'eron, D. Esteve, and M. H. Devoret, Phys. Rev. Lett.
	{\bf 73}, 2488 (1994). 

\bibitem{magnee}  P. H. C. Magnee, N. van der Post, P. H. M. Kooistra, 
B. J. van Wees,
	and T. M. Klapwijk, Phys. Rev. B {\bf 50}, 4594 (1994).

\bibitem{tech}  J. R. Gao, J. P. Heida, B. J. van Wees, S. Bakker, and T.
	M. Klapwijk, Appl. Phys. Lett. {\bf 63}, 334 (1993); 
        T. Akazaki, J. Nitta, and H. Takayanagi, Appl. Phys.
	Lett. {\bf 59}, 2037 (1991); 
	R. Taboryski, T. Clausen, J. Bindslev Hansen, J. L.
	Skov, J. Kutchinsky, C.B. S{\o}rensen, and P. E. Lindelof, Appl. Phys.
	Lett. {\bf 69}, 656 (1996); 
	S. De Franceschi, F. Giazotto, F. Beltram, L. Sorba, M. Lazzarino,
	and A. Franciosi, Appl. Phys. Lett. {\bf 73}, 3890 (1998).

\bibitem{neur} K. Neurohr, A. A. Golubov, Th. Klocke, J. Kaufmann, 
Th. Sch\"{a}pers,
	J. Appenzeller, D. Uhlisch, A. V. Ustinov, M. Hollfelder, 
H. L\"{u}th, and
	A. I. Braginski, Phys. Rev. B {\bf 54}, 17018 (1996).

\bibitem{ldos}
        S. Gu\'eron, H. Pothier, N. O. Birge, D. Esteve,
        and M. Devoret, Phys. Rev. Lett. {\bf 77}, 3025 (1996);
        A. A. Golubov and M. Yu. Kupriyanov, J. Low Temp.
        Phys. {\bf 70}, 83 (1988);
        W. Belzig, C. Bruder, and G. Sch\"on, Phys. Rev. B {\bf 54},
        9443 (1996).

\bibitem{schmid81}A. Schmid in {\em Nonequilibrium
        Superconductivity, Phonons, and Kapitza Boundaries},
        NATO ASI Series B 65,
        ed. K.E. Gray, (Plenum, New York 1981).

\bibitem{vaglio} A. Andreone, A. Cassinese, M. Iavarone, R. Vaglio, 
I. I. Kulik,
	and V. Palmieri, Phys. Rev. B {\bf 52}, 4473 (1995).


\bibitem{z} In the framework of Ref.\onlinecite{benPrl}, 
the ratio $r$ is equal to 
$(\Gamma{d/l_e})^{-1}$, where $\Gamma$ is the tunnel-barrier transmission 
probability. This allows us to estimate $\Gamma\simeq 0.44$.
In the language of the widespread ballistic model of 
G. E. Blonder, M. Tinkham, and T. M. Klapwijk, Phys. Rev. 
B {\bf 25}, 4515 (1982), the latter value translates into
a barrier strength parameter $Z\simeq 1.1$.  
In contrast, a straightforward analysis within the ballistic model would give 
$\Gamma=0.55$ (i.e. $Z=0.9$). This overestimate 
of the interface transparency stems largely from neglecting subgap 
single-particle current. 

\bibitem{area} W. M. van Huffelen, T. M. Klapwijk, 
D.R. Heslinga, M. J. Boer, and N. van der Post, Phys. Rev. 
B {\bf 47}, 5170 (1993). 

\end{references}
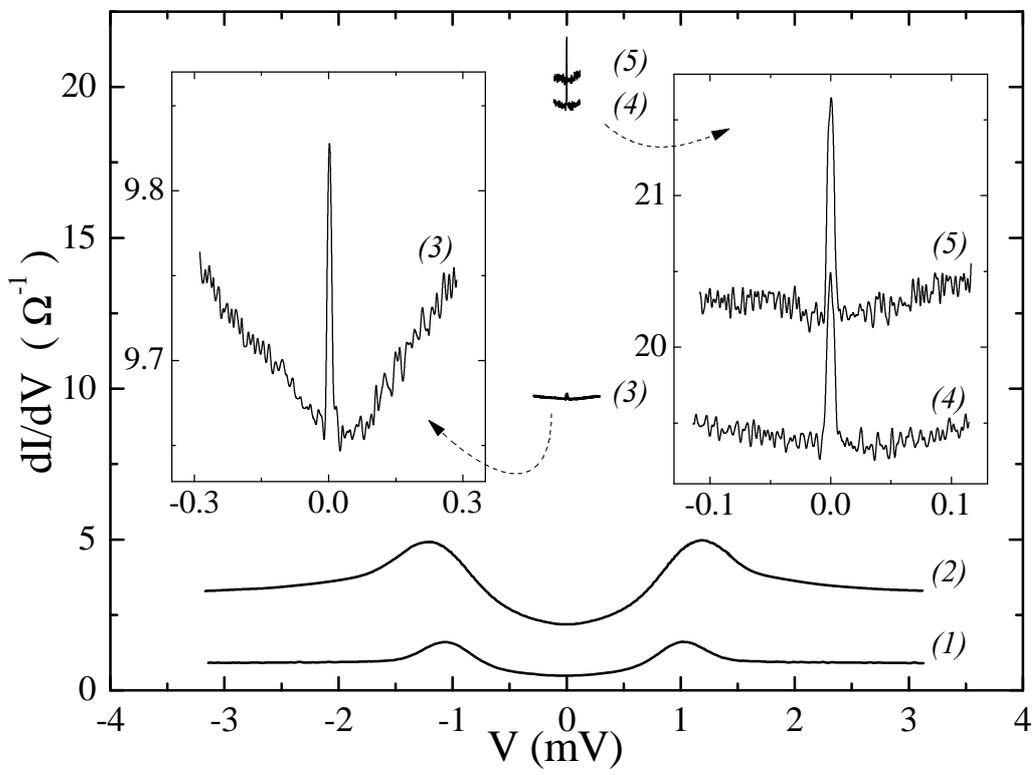
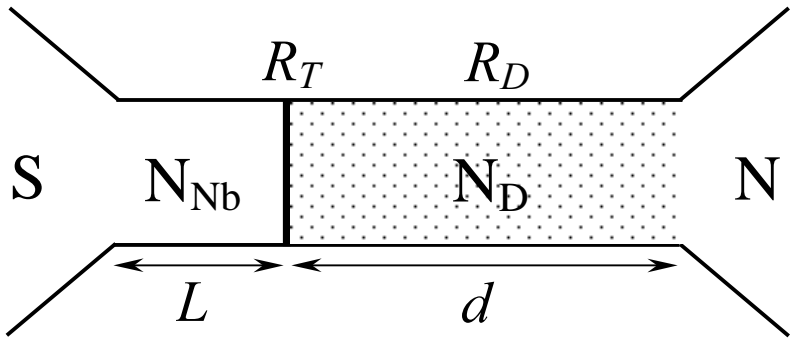
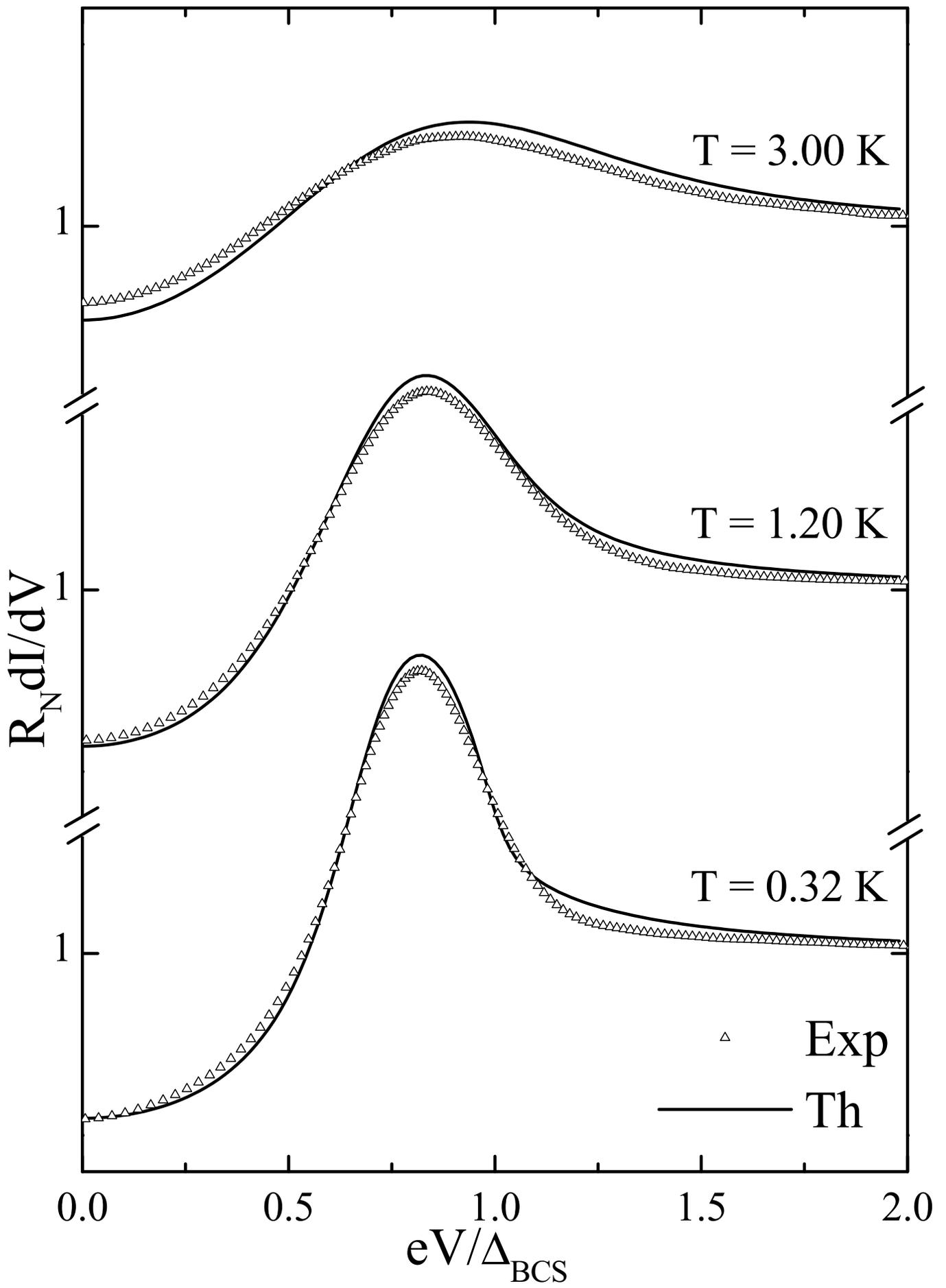
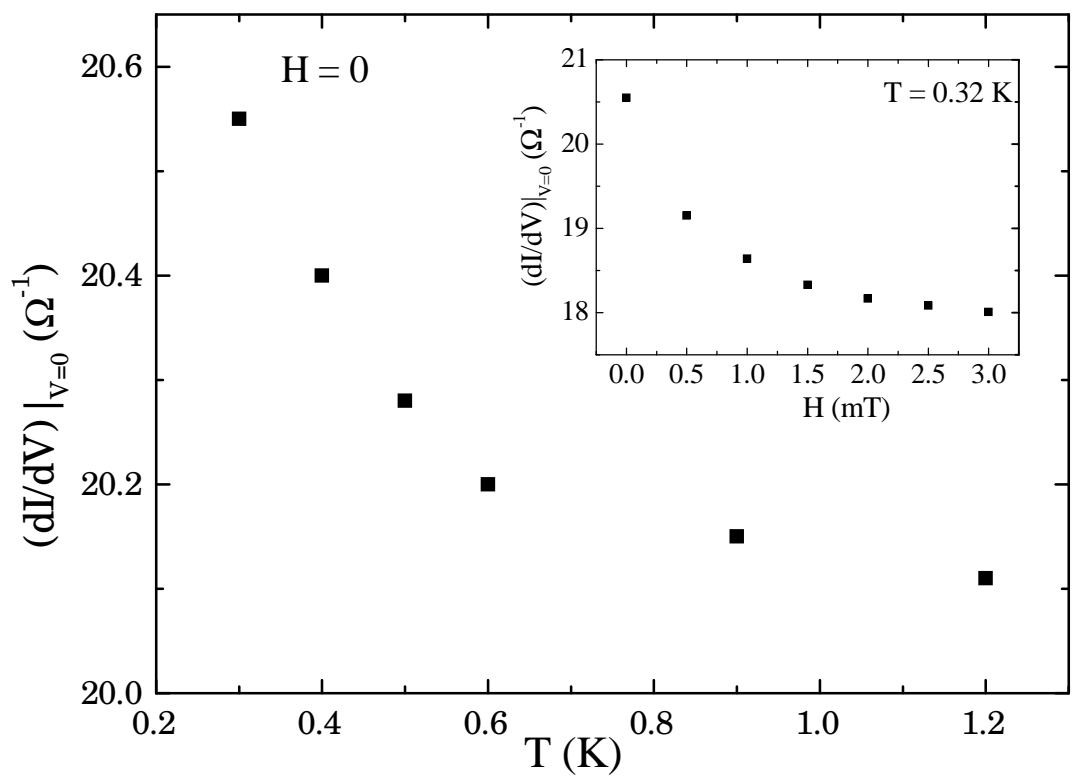
\end{document}